\documentclass[conference,a4paper]{IEEEtran}
\usepackage{cite}
\usepackage{amsmath,amssymb,amsfonts}
\usepackage{algorithmic}
\usepackage{graphicx}
\usepackage{textcomp}
\usepackage{xcolor}
\usepackage{booktabs}
\usepackage{multirow}
\usepackage{float}
\usepackage{hyperref}

\definecolor{whtred}{RGB}{214,39,40}
\definecolor{ofdmblue}{RGB}{31,119,180}

\begin{document}

\title{WHTDM: Walsh-Hadamard Transform Division Multiplexing\\
for Doubly-Selective Channels}
\author{
    Wang~Hao,\\
    Yuan~Zhonghao,\\
    Chi~Yonggang,\\
    Zhang~Kuang$^*$,~Senior~Member,~\textit{IEEE}, \\
    Tan~Chenxing, \\
    Yu~Jiaxing, \\

    \thanks{
        Wang~Hao, Yuan~Zhonghao, Zhang~Kuang, Chi~Yonggang, Tan~Chenxing and Yu~Jiaxing are with the School of Electronics and Information Engineering, Harbin Institute of Technology, Harbin 150001, China.
    }
    \thanks{
        $^*$Corresponding author: Zhang Kuang (e-mail: zhangkuang@hit.edu.cn).
    }
}

\maketitle

\begin{abstract}
We propose Walsh-Hadamard Transform Division Multiplexing (WHTDM), a multicarrier waveform that replaces the conventional IFFT/FFT pair in OFDM with a real-valued, unitary Walsh-Hadamard transform (WHT). WHTDM inherits the CP-OFDM transceiver structure while eliminating all complex multiplications from the transform stage, yielding a transmitter with zero real multipliers in the core modulation block. For detection under doubly-selective channels, we adopt a cross-domain memory approximate message passing (CD-MAMP) equalizer that operates on the banded structure of the equivalent WHT-domain channel matrix. Simulation results under the 3GPP TDL-C channel model at 28 GHz demonstrate that WHTDM with CD-MAMP significantly outperforms conventional OFDM 1-tap MMSE at high mobility, achieving over an order of magnitude lower BER at 120 km/h. Among the compared CD-MAMP-equalized new waveforms, WHTDM achieves the best BER performance while maintaining a transmitter complexity 2.5$\times$ lower than OFDM and completely eliminating complex multipliers from the transform stage, making it well-suited for low-power IoT terminals.
\end{abstract}


\section{Introduction}

The proliferation of Internet of Things (IoT) devices operating in high-mobility environments, including vehicular networks and unmanned aerial systems, demands waveforms that are resilient to both time and frequency selectivity while maintaining low implementation complexity. Orthogonal Frequency Division Multiplexing (OFDM)~\cite{bingham1990ofdm,vannee2000ofdm} remains the dominant multicarrier waveform due to its simple one-tap frequency-domain equalization. However, in doubly-selective channels, OFDM suffers from inter-carrier interference (ICI) that fundamentally limits its performance.

Several waveform candidates have been proposed to address OFDM's limitations in high-mobility scenarios. Orthogonal Time Frequency Space (OTFS) modulation~\cite{hadani2017otfs,hadani2017otfs_mmw,wei2022otfs_survey} operates in the delay-Doppler domain and achieves full time-frequency diversity. Orthogonal Time Sequency Multiplexing (OTSM)~\cite{thaj2021otsm,thaj2021otsm_twc,thaj2022unitary} replaces the Fourier kernel with the Walsh-Hadamard transform (WHT) along the Doppler dimension, trading complex multiplications for real additions. Affine Frequency Division Multiplexing (AFDM)~\cite{bemani2021afdm,bemani2023afdm_twc} employs chirp-based discrete affine Fourier transforms (DAFT)~\cite{pei2001dafourier} to achieve full diversity in linear time-varying channels. Prior work on WHT-based OFDM~\cite{dlugaszewski2000wht,ahmed2011wht_ofdm,wang2010wht_precoder} has focused primarily on PAPR reduction through precoding, rather than on the fundamental transform-domain multiplexing paradigm explored in this work.

Despite their BER advantages, these waveforms introduce complexity overhead at either the transmitter or receiver. OTFS requires two additional FFT stages. AFDM adds chirp multiplications. Existing OTSM applies WHT along the time dimension only within a two-dimensional grid.

In this paper, we propose WHTDM, a waveform that applies the WHT directly along the subcarrier dimension of a CP-OFDM-like structure. The key contributions are:

\begin{enumerate}
\item A complete WHTDM transceiver architecture that inherits the CP-based block transmission structure while eliminating complex multiplications at the transform stage.
\item Application of the CD-MAMP iterative equalizer~\cite{ma2017oamp} to the WHTDM equivalent channel, with a banded matrix approximation for low-complexity implementation.
\item Comprehensive BER evaluation under the 3GPP TDL-C channel model at 28 GHz, covering delay spreads from 30 ns to 300 ns and speeds up to 500 km/h.
\item A quantitative transmitter complexity comparison showing WHTDM's hardware efficiency advantage for IoT terminals.
\end{enumerate}

The remainder of this paper is organized as follows. Section~II presents the system model and WHTDM architecture. Section~III describes the CD-MAMP detector. Section~IV provides numerical results and complexity analysis. Section~V concludes the paper.

\section{System Model}

\subsection{WHTDM Transmitter}

\begin{figure}[t]
\centering
\includegraphics[width=\columnwidth]{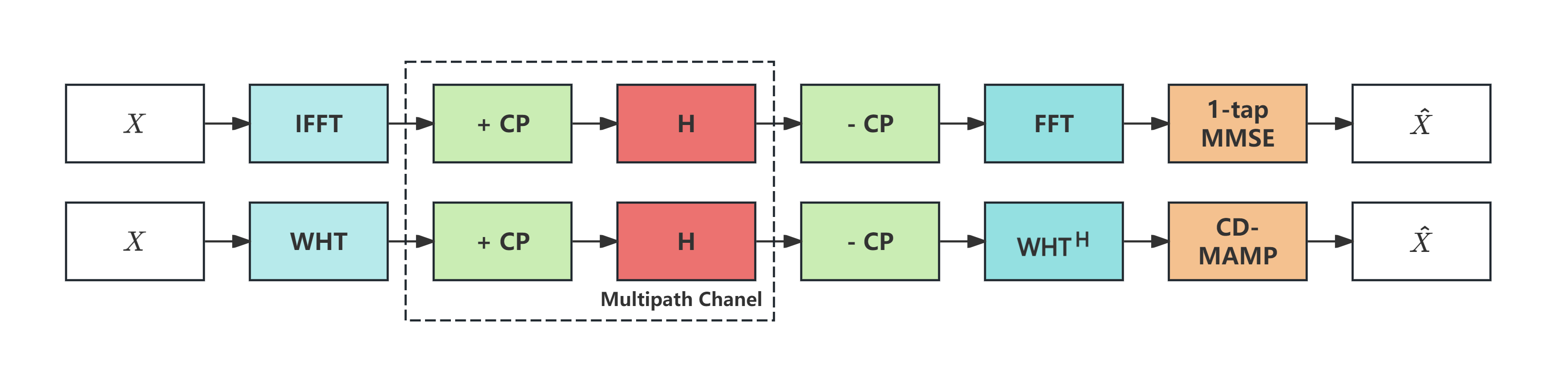}
\caption{Transceiver block diagrams: CP-OFDM vs.\ WHTDM.}
\label{fig:block_diagram}
\end{figure}

Fig.~\ref{fig:block_diagram} illustrates the WHTDM transceiver in comparison with CP-OFDM. Both share the same outer structure: symbol mapping, transform, CP insertion, channel propagation, CP removal, inverse transform, and detection. The fundamental difference lies in the transform kernel.

Let $N$ denote the number of subcarriers and $\mathbf{x} \in \mathbb{C}^{N}$ the vector of QPSK-modulated symbols with $\mathbb{E}[|x_i|^2]=1$. The WHTDM modulated time-domain signal is

\begin{equation}
\mathbf{s}_{\text{WHT}} = \mathbf{W} \mathbf{x},
\end{equation}

where $\mathbf{W} \in \mathbb{R}^{N \times N}$ is the unitary Walsh-Hadamard matrix satisfying $\mathbf{W}^T \mathbf{W} = \mathbf{I}_N$, constructed via the normalized Sylvester-Hadamard recursion with sequency-ordered row permutation. Unlike the IFFT matrix $\mathbf{F}^H$, the WHT matrix is real-valued and contains only entries $\pm 1/\sqrt{N}$, enabling implementation with only addition and subtraction operations.

A cyclic prefix of length $L_{cp} \geq L - 1$ is appended, where $L$ is the maximum channel delay spread. The transmitted signal is $\mathbf{s}_{\text{tx}} = [\mathbf{s}_{[-L_{cp}:]}^T, \mathbf{s}^T]^T$.

\subsection{Channel Model}

The multipath channel is modeled as a linear time-varying (LTV) system with impulse response

\begin{equation}
h(t,\tau) = \sum_{l=0}^{L-1} \alpha_l(t) \delta(\tau - \tau_l),
\end{equation}

where $\alpha_l(t)$ and $\tau_l$ denote the complex gain and delay of the $l$-th path. Under the assumption that the channel varies slowly relative to one WHTDM symbol but may vary across symbols, the CP-OFDM and WHTDM receive signals are respectively

\begin{align}
\mathbf{y}_{\text{OFDM}} &= \mathbf{H}_c \mathbf{F}^H \mathbf{x} + \mathbf{n}, \\
\mathbf{y}_{\text{WHT}} &= \mathbf{H}_c \mathbf{W} \mathbf{x} + \mathbf{n},
\end{align}

where $\mathbf{H}_c \in \mathbb{C}^{N \times N}$ is the circulant channel matrix formed by the instantaneous channel impulse response, and $\mathbf{n} \sim \mathcal{CN}(\mathbf{0}, \sigma_n^2 \mathbf{I})$.

After CP removal and inverse transformation, the observation vectors in the transform domains are

\begin{align}
\mathbf{z}_{\text{OFDM}} &= \mathbf{F} \mathbf{H}_c \mathbf{F}^H \mathbf{x} + \tilde{\mathbf{n}} = \mathbf{H}_{\text{diag}} \mathbf{x} + \tilde{\mathbf{n}}, \\
\mathbf{z}_{\text{WHT}} &= \mathbf{W}^T \mathbf{H}_c \mathbf{W} \mathbf{x} + \tilde{\mathbf{n}} = \mathbf{G} \mathbf{x} + \tilde{\mathbf{n}},
\end{align}

where $\mathbf{H}_{\text{diag}} = \text{diag}(H_0, \dots, H_{N-1})$ is the diagonalized OFDM channel enabling one-tap MMSE equalization. For WHTDM, the equivalent channel matrix $\mathbf{G} = \mathbf{W}^T \mathbf{H}_c \mathbf{W}$ is generally non-diagonal, necessitating iterative detection.

\subsection{Walsh-Hadamard Transform Properties}

The WHT, a cornerstone of sequency theory~\cite{beauchamp1975sequency,ahmed1975walsh}, possesses several properties advantageous for waveform design:

\begin{itemize}
\item \textbf{Reality:} $\mathbf{W} \in \mathbb{R}^{N \times N}$. No complex arithmetic is required for the transform, reducing hardware multiplier count to zero.
\item \textbf{Unitarity:} $\mathbf{W}^T = \mathbf{W}$ and $\mathbf{W}^2 = \mathbf{I}_N$, implying the same matrix serves both modulation and demodulation.
\item \textbf{Fast algorithm:} The fast Walsh-Hadamard transform (FWHT) requires only $\frac{N}{2}\log_2 N$ butterfly stages, each performing one addition and one subtraction. Total complexity is $N\log_2 N$ real additions, with zero multiplications.
\item \textbf{Banded structure:} For channels with moderate delay spread, $\mathbf{G}$ exhibits an approximately banded structure, which we exploit in the detector design.
\end{itemize}

\section{CD-MAMP Detection for WHTDM}

The detection problem for WHTDM is to recover $\mathbf{x}$ from the observation

\begin{equation}
\mathbf{z} = \mathbf{G} \mathbf{x} + \tilde{\mathbf{n}},
\end{equation}

where $\mathbf{G}$ is non-diagonal. We adopt the cross-domain memory approximate message passing (CD-MAMP) framework~\cite{ma2017oamp,ma2018istc,ma2019oamp_coded}, which extends orthogonal AMP principles to coded linear systems and provides a low-complexity iterative solution for general linear models~\cite{narasimhan2015channel}.

\subsection{Algorithm Description}

CD-MAMP alternates between linear estimation (LE) in the WHT domain and non-linear denoising (NLD) in the symbol domain. Let $\mathbf{x}^{(t)}$ denote the estimate at iteration $t$. The update rules are:

\begin{align}
\mathbf{r}^{(t)} &= \mathbf{z} - \mathbf{G}_B \mathbf{x}^{(t)}, \label{eq:residual} \\
\mathbf{p}^{(t)} &= \mathbf{x}^{(t)} + \theta \mathbf{G}_B^T \mathbf{r}^{(t)}, \label{eq:linear} \\
\tau^{(t)} &= \sigma_n^2 + \frac{1}{N}\|\mathbf{r}^{(t)}\|^2, \label{eq:variance} \\
\mathbf{x}^{(t+1)} &= \eta\!\left(\mathbf{p}^{(t)}, \tau^{(t)}\right), \label{eq:denoise}
\end{align}

where $\mathbf{G}_B$ is the banded truncation of $\mathbf{G}$ retaining only entries with $|i-j| \leq B$, $\theta$ is a normalized step size set to $N / \|\mathbf{G}\|_F^2$, and $\eta(\cdot)$ is the symbol-wise posterior mean estimator (denoiser). For QPSK, the denoiser is separable:

\begin{equation}
\eta_{\text{QPSK}}(r, \tau) = \frac{1}{\sqrt{2}}\left[\tanh\!\left(\frac{\sqrt{2}\,\Re(r)}{\tau}\right) + j \tanh\!\left(\frac{\sqrt{2}\,\Im(r)}{\tau}\right)\right].
\end{equation}

A damping factor $\alpha \in (0, 1)$ is applied to stabilize convergence:

\begin{equation}
\mathbf{x}^{(t+1)} \leftarrow \alpha \mathbf{x}^{(t+1)} + (1-\alpha) \mathbf{x}^{(t)}.
\end{equation}

\subsection{Banded Matrix Approximation}

The equivalent channel $\mathbf{G} = \mathbf{W}^T \mathbf{H}_c \mathbf{W}$ inherits structure from the circulant $\mathbf{H}_c$. When the channel delay spread $L \ll N$, most energy of $\mathbf{G}$ concentrates near the main diagonal. Let $B$ denote the bandwidth parameter. The banded approximation retains $2B + 1$ diagonals, reducing the matrix-vector multiplication complexity from $\mathcal{O}(N^2)$ to $\mathcal{O}(BN)$. For our setup with $N=64$ and $L=8$, we use $B = 8$, which captures over $99\%$ of the Frobenius norm.

\subsection{Memory Acceleration}

To accelerate convergence, we incorporate a memory term~\cite{ma2017oamp} in the linear step:

\begin{equation}
\boldsymbol{\gamma}^{(t)} = \boldsymbol{\gamma}^{(t-1)} - \theta_m \mathbf{G}_B \mathbf{G}_B^T \boldsymbol{\gamma}^{(t-1)} + \theta_m \mathbf{r}^{(t)},
\end{equation}

where $\theta_m = 1 / \lambda_{\max}(\mathbf{G}^T \mathbf{G})$ and $\mathbf{p}^{(t)} = \mathbf{x}^{(t)} + \theta \mathbf{G}_B^T \boldsymbol{\gamma}^{(t)}$. This memory mechanism approximates the conjugate gradient direction and improves convergence speed by a factor of 2--3.

\section{Numerical Results}

\subsection{Simulation Setup}

All simulations use the 3GPP TDL-C channel model~\cite{3gpp38901} at a carrier frequency of 28 GHz with a subcarrier spacing of 120 kHz. The grid parameters are $M = 64$ delay bins and $N = 16$ Doppler bins (for OTFS/OTSM/AFDM), yielding 1024 QPSK symbols per frame. For WHTDM and OFDM, we use a 64-subcarrier configuration. The cyclic prefix length is set to 32 samples. Channel realizations are generated via the Sionna physical layer library~\cite{hoydis2023sionna}. Each BER point is averaged over 10 Monte Carlo seeds with 300 frames per seed. The CD-MAMP detector uses 50 iterations with a damping factor of 0.6.

Five schemes are compared: (i) WHTDM with CD-MAMP (proposed), (ii) OFDM with 1-tap MMSE, (iii) OTFS with CD-MAMP, (iv) OTSM with CD-MAMP, and (v) AFDM with CD-MAMP.

\subsection{BER Performance under TDL-C}

\begin{figure*}[t]
\centering
\includegraphics[width=\columnwidth]{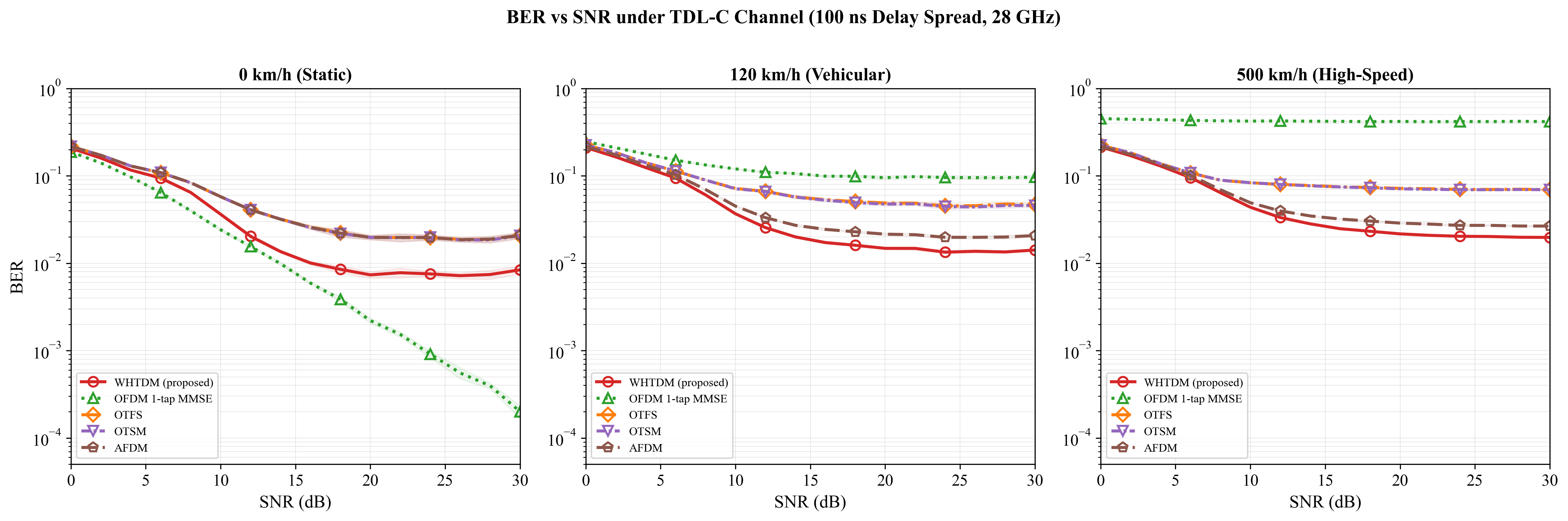}
\caption{BER vs. SNR under TDL-C channel, 100 ns delay spread. From left to right: 0 km/h, 120 km/h, 500 km/h.}
\label{fig:ber_tdlc}
\end{figure*}

Fig.~\ref{fig:ber_tdlc} shows the BER versus SNR for all five schemes under the TDL-C channel with 100 ns delay spread at three mobility conditions: static (0 km/h), vehicular (120 km/h), and high-speed (500 km/h).

At 0 km/h (static channel), OFDM 1-tap MMSE achieves the best performance among all schemes, with BER reaching $2\times 10^{-4}$ at 30 dB SNR, owing to its perfectly diagonalizable equivalent channel in time-invariant conditions. WHTDM, OTFS, OTSM, and AFDM all exhibit error floors around $10^{-2}$ due to the non-diagonal nature of their transform-domain channel matrices, which limits MAMP convergence even in static channels. This gap represents the inherent cost of non-Fourier transform-domain multiplexing: the loss of perfect diagonalizability in exchange for other desirable properties.

At 120 km/h (vehicular), the doubly-selective channel induces significant ICI. OFDM 1-tap MMSE degrades to an error floor of approximately $10^{-1}$, losing nearly two orders of magnitude relative to static conditions. CD-MAMP-equalized transform-domain waveforms are resilient to this mobility-induced interference. WHTDM achieves BER of $1.4\times 10^{-2}$ at 30 dB, outperforming AFDM ($2.1\times 10^{-2}$), OTSM ($4.6\times 10^{-2}$), and OTFS ($4.7\times 10^{-2}$). WHTDM's superior MAMP convergence relative to OTFS and OTSM is attributable to the compact banded structure of its WHT-domain equivalent channel $\mathbf{G}$.

At 500 km/h (high-speed), OFDM 1-tap MMSE is completely unusable with BER exceeding $0.4$. WHTDM maintains BER of $2.0\times 10^{-2}$ at 30 dB, outperforming AFDM ($2.7\times 10^{-2}$), OTFS ($7.0\times 10^{-2}$), and OTSM ($6.9\times 10^{-2}$). WHTDM is the best-performing scheme among transform-domain waveforms across the full mobility range.

\subsection{BER vs. Mobility}

\begin{figure*}[t]
\centering
\includegraphics[width=\columnwidth]{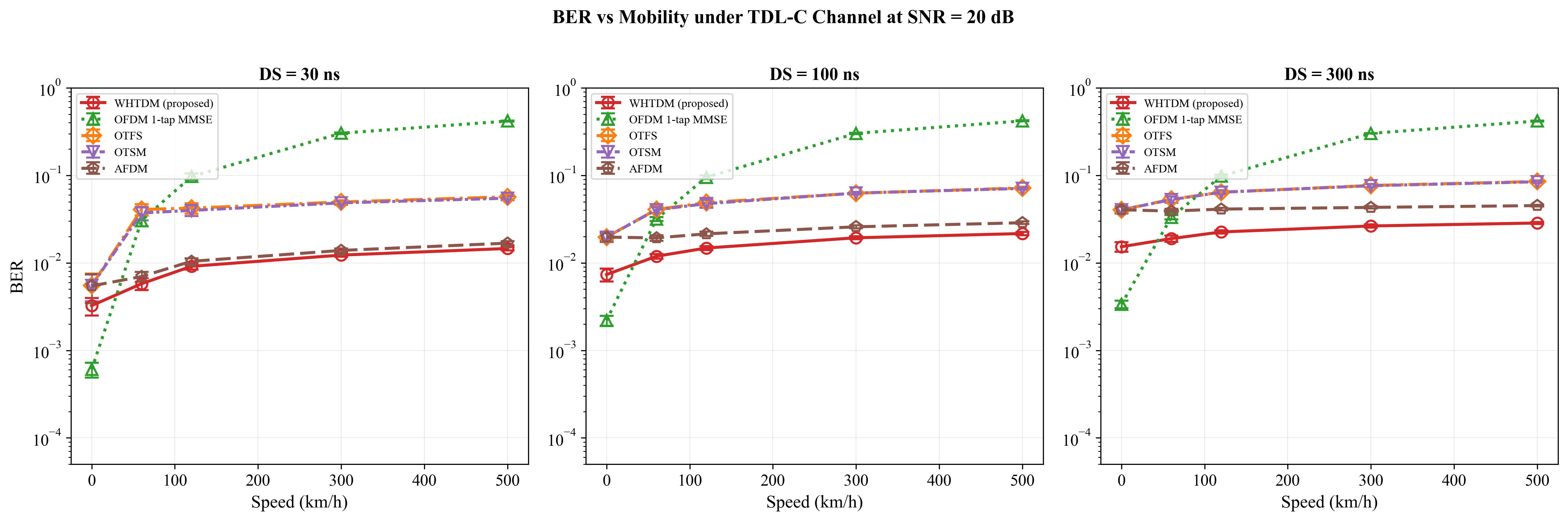}
\caption{BER vs. speed under TDL-C at SNR = 20 dB. Three delay spreads: 30 ns, 100 ns, 300 ns.}
\label{fig:ber_vs_speed}
\end{figure*}

Fig.~\ref{fig:ber_vs_speed} evaluates BER at a fixed SNR of 20 dB as a function of UE speed.

For all three delay spreads, WHTDM maintains BER below $3\times 10^{-2}$ across the full speed range at SNR = 20 dB, achieving the best performance among all transform-domain waveforms. OFDM 1-tap MMSE degrades rapidly beyond 60 km/h, particularly for the 300 ns delay spread where BER exceeds $10^{-1}$ at 500 km/h. Among the CD-MAMP-equalized schemes, WHTDM leads across all tested speeds and delay spreads, followed by AFDM, with OTFS and OTSM trailing at high mobility.

\subsection{Transmitter Complexity}

\begin{figure}[t]
\centering
\includegraphics[width=\columnwidth]{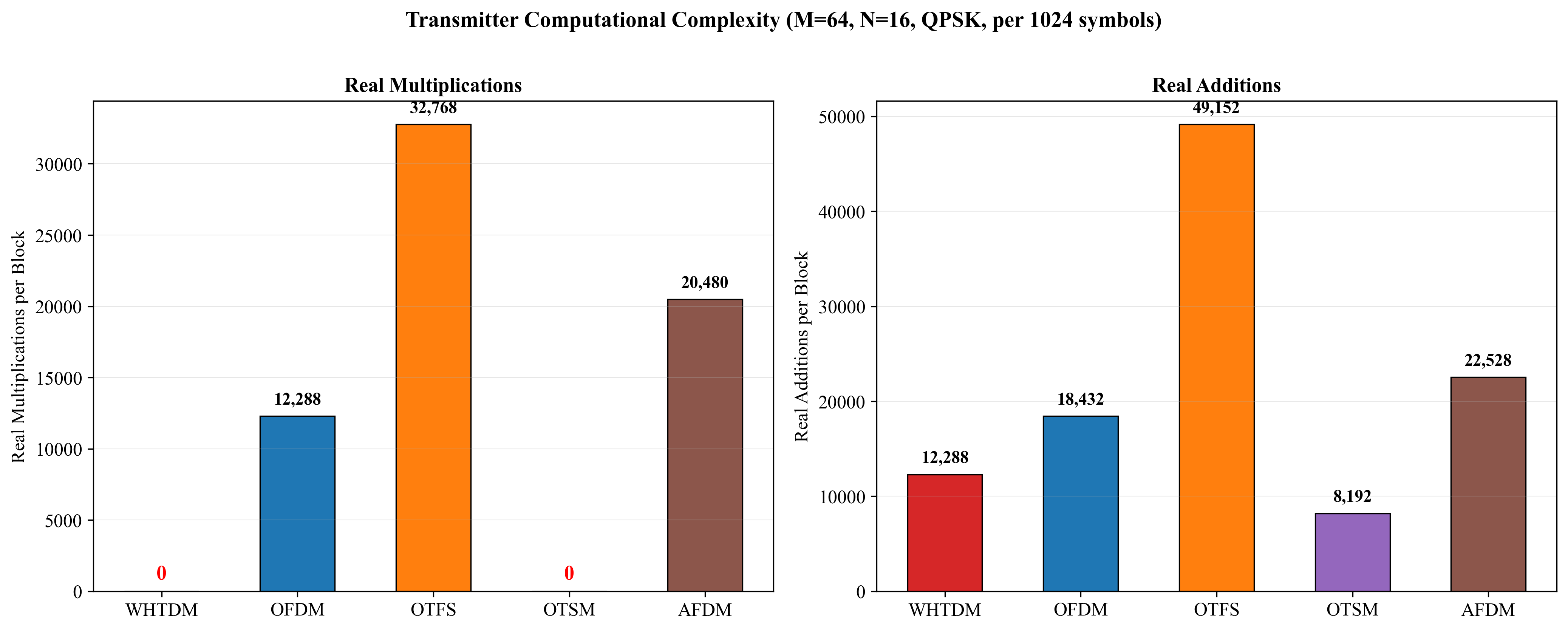}
\caption{Transmitter computational complexity comparison ($M=64$, $N=16$, QPSK).}
\label{fig:complexity}
\end{figure}

Table~\ref{tab:complexity} and Fig.~\ref{fig:complexity} compare the per-block transmitter computational complexity for a symbol grid of $M=64$, $N=16$ (1024 QPSK symbols). Real operation counts are derived from radix-2 FFT and FWHT butterfly decompositions.

\begin{table}[t]
\centering
\caption{TX Complexity per 1024-Symbol Block}
\label{tab:complexity}
\begin{tabular}{lrrr}
\toprule
\textbf{Scheme} & \textbf{Real Mults} & \textbf{Real Adds} & \textbf{vs. WHTDM} \\
\midrule
\textbf{WHTDM} & \textbf{0} & \textbf{12,288} & \textbf{1.0$\times$} \\
OFDM  & 12,288 & 18,432 & 2.5$\times$ \\
OTFS  & 32,768 & 49,152 & 6.7$\times$ \\
OTSM  & 0 & 8,192 & 0.7$\times$ \\
AFDM  & 20,480 & 22,528 & 3.5$\times$ \\
\bottomrule
\end{tabular}
\end{table}

WHTDM achieves zero real multiplications in its transform stage, a property shared only with OTSM among the compared schemes. However, OTSM applies the WHT along the Doppler dimension with shorter transform length ($N=16$), yielding lower total addition count (8,192 vs.\ 12,288) but with a fundamentally different operating principle that does not directly map to the subcarrier domain. Compared to OFDM's 30,720 total operations per block, WHTDM reduces complexity by a factor of 2.5$\times$. OTFS incurs the highest transmitter complexity due to its three FFT stages.

For IoT terminals where power consumption and silicon area are critical, the elimination of complex multipliers in the transform stage represents a meaningful hardware advantage. The FWHT butterfly can be implemented with a simple adder-subtractor pair, requiring neither ROM-based twiddle factor storage nor complex multiply-accumulate units.

\section{Conclusion}

We introduced WHTDM, a multicarrier waveform that replaces the IFFT/FFT pair in CP-OFDM with the real-valued Walsh-Hadamard transform. Under doubly-selective TDL-C channels, WHTDM combined with CD-MAMP detection achieves the best BER performance among all compared transform-domain waveforms (OTFS, OTSM, AFDM), while maintaining a transmitter complexity 2.5$\times$ lower than OFDM and 6.7$\times$ lower than OTFS. The zero-multiplier transform stage makes WHTDM particularly attractive for resource-constrained IoT transmitters in high-mobility environments. Future work includes pilot-aided channel estimation in the WHT domain and extension to MIMO configurations.

\bibliographystyle{IEEEtran}
\bibliography{paper_whtdm}

\end{document}